\documentclass{elsarticle}

\usepackage{amsmath} 
\usepackage{graphicx,psfrag,color}

\usepackage{upgreek}

\usepackage{amsmath}
\usepackage{amsfonts}
\usepackage{calrsfs}
\usepackage{mathrsfs}

\newcommand{\su}{\mathtt{u}}

\DeclareMathOperator{\totwave}{t}
\DeclareMathOperator{\inc}{\text{in}}
\DeclareMathOperator{\sconst}{\mathtt{C}}

\usepackage[colorlinks, citecolor=blue, linkcolor=blue, backref=true]{hyperref}

\journal{International Journal of Engineering Science}

\begin{document}

\begin{frontmatter}

\title{Wave transmission across surface interfaces in lattice structures}

 \author[IITK]{Basant Lal Sharma}
 \ead{bls@iitk.ac.in}
\author[PG,sfedu]{Victor A. Eremeyev}
 \ead{eremeyev.victor@gmail.com}
\address[IITK]{Department of Mechanical Engineering, Indian Institute of Technology Kanpur, Kanpur 208016, India}
 \address[PG]{Faculty of Civil and Environmental Engineering, Gda\'{n}sk University of Technology, \\ ul. Gabriela Narutowicza 11/12
 80-233 Gda\'{n}sk, Poland}
\address[sfedu]{R. E. Alekseev Nizhny Novgorod Technical University,  Minin St., 24, Nizhny Novgorod, 603950 Russia}

\begin{abstract}
 Within the lattice dynamics formulation, we present an exact solution for anti-plane surface waves in a square lattice strip with a surface row of material particles of two  types separated by a linear interface. The considered problem is a discrete analog of an elastic half-space with surface stresses modelled through the simplified Gurtin--Murdoch model, where we have an interfacial line separating areas with different surface elastic properties. The main attention is paid to the transmittance and the reflectance of a wave across the interface. The presented results shed a light on the influence on surface waves of surface inhomogeneity in surface elastic properties such as grain and subgrain boundaries.
\end{abstract}


\end{frontmatter}

\section*{Introduction}

Nowadays the interest in application of the surface elasticity models keeps growing with recent achievements in the nanotechnology, see, e.g., \cite{Duanetal2008,Wang2011,javili2013thermomechanics,javili2013geometrically,eremeyev2016effective} and the reference therein. In particular, these models can forecast size-effect observed at the nanoscale and other phenomena related to high ratio of material particles in the vicinity of a surface/interface to ones in the bulk.  Having an origin in the seminal works by \cite{Laplace1805a,Laplace1805b,Young1806,Poisson1831} and Gibbs, see \cite{Gibbs1928}, within the   finite elasticity the first model of surface elasticity was proposed by \cite{GurtinMurdoch1975a,gurtin1978surface}. The proposed model describes finite deformations of an elastic solid body with an elastic membrane  glued on its surface. The further extension of the Gurtin--Murdoch model was proposed by  \cite{steigmann1997plane,SteigmannOgden1999} who considered also a bending stiffness of a surface structure. Some further extensions of the surface elasticity were proposed by \cite{Placidi01072014,LURIE2009709,anatolyevich2019classifying,EremeyevRSTA2019}.

The presence of surface microstructure  results in changes of effective properties of composites, see, \textit{e.g.}, \citep{sevostianov2007effect,kushch2013elastic,zemlyanova2017circular,HAN2018166}  and the reference therein, or even in appearance of new phenomena such as the existence of anti-plane surface waves \cite{xu2015anti,eremeyev2016surface}. It is worth reminding that surface acoustic waves are widely used in the nondestructive strength evaluation as they may bring some information on material structure in the vicinity of a surface, see, e.g., \cite{Uberall1973wl,achenbach2012wave}. Thus, the further analysis of models based on the surface elasticity approach lies in the focus of current researches.

For all continuum models of surface elasticity one should introduce constitutive relations on the surface additionally to ones defined in the bulk. This axiomatic way of the problem statement found some confirmation through discrete models, such the lattice or molecular dynamics described by \cite{brillouin1946wave,BornHuang1985,hoover1986molecular,Maradudin}. In particular, the latter results in reasonable values of surface elastic parameters, see, \textit{e.g.},  \cite{miller2000size,shenoy2005atomistic}. In the same time discrete models can describe in all details the material behaviour in the vicinity of crack tips, changes in geometry and other geometrical singularities, see, \textit{e.g.} \cite{mishuris2007waves,mishuris2009localised,slepyan2012models,sharma2015diffraction,sharma2017scattering,porubov2013nonlinear,porubov2018two,gorbushin2018influence}, where the efficiency of discrete lattice models is demonstrated. As an example, we recently provided the comparison of the surface wave propagation within the Gurtin--Murdoch model and the lattice dynamics which shows similarities in dispersion relations for both models after some re-scaling, see \cite{Victor_Bls_surf1}. Thus, the lattice dynamics can serve as a counterpart of   continuum surface elasticity models.

Let us note that almost all problems considered within the surface elasticity relate to constant surface properties. An exceptional case presented by \cite{wang2015two,wang2016mode} demonstrated show that inhomogeneity in surface elastic properties may significantly affect the corresponding solutions. 

\begin{figure}[ht!]
\center{\includegraphics[width=.8\textwidth]{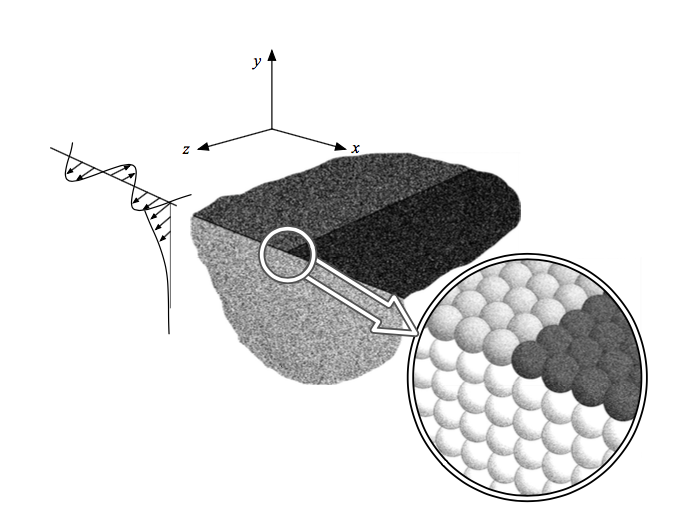}}
\caption[ ]{Geometry of the three dimensional lattice half space is shown schematically by placing a cutting section along $x$-$y$ plane. On the right side (by zooming into the central portion), an atomic/particle based discrete model of an elastic half space with surface interface \cite{Victor_Bls_surf1} is shown. The kinematics assumed is that if anti-plane motion, i.e. the displacement occurs along $z$ direction only, as shown on the left, while it is independent of the $z$ coordinate.}
\label{Fig1}
\end{figure}

Following \cite{Victor_Bls_surf1} here we discuss the anti-plane surface waves propagation in an elastic half-space with a piece-wise inhomogeneity in surface properties. We consider a square lattice which consists of material particles of mass $M$ and stiffness of bonds $K$, where the surface consists of material particles of two kind which both are different from the ones in the bulk. The surface material particles are separated by a line which constitute a surface interface. In other words we consider a discrete model of an elastic halfspace ${\mathtt{y}}\le 0$ with surface stresses defined at ${\mathtt{y}}=0$ such that the  surface elastic moduli and surface mass density are constant but different in half-planes ${\mathtt{x}}>0$ and ${\mathtt{x}}<0$, respectively, see Fig.~\ref{Fig1}. The aim of the paper is to analyze the influence of the surface  interface on the wave propagation.

Suppose ${\mathscr{R}}_{S}$ and ${\mathscr{T}}_{S}$ denote the surface wave reflectance and transmittance \citep{brillouin1946wave} relative to the surface interface and given incident surface wave. These entities are respectively defined as the reflected and transmitted energy flux in the surface wave per unit energy flux of the incident surface wave. Note that ${\mathscr{T}}_{S}=0$ in the absence of a surface structure on the other side of the interface, in which case even the entity ${\mathscr{R}}_{S}$ is an interesting entity to discover behaviour relative to the incident wave parameters and material parameters on the side where the surface wave is generated.
In general, there is certain amount of energy lost into the bulk during the process of transmission on the surface. This is denoted by ${\mathscr{D}}_{S}=1-{\mathscr{R}}_{S}-{\mathscr{T}}_{S}$. Indeed, due to energy conservation, the energy flux ${\mathscr{D}}_{S}$ is carried by all other wave modes in the two sides of the surface interface (besides the two surface wave modes). It is a fundamental question how ${\mathscr{D}}_{S}$ depends on the wave number of the surface wave as well as material parameters associated with the surface structure on both sides. The question is answered in the present paper and an exact expression is provided. Moreover, an additional question becomes an interesting follow up of this paper concerning the continuum limit: to elastic half space following the work of \cite{Victor_Bls_surf1,BlsNearCrack,Bls31} as well as the limit to an elastic strip; this question has been deferred to another paper elsewhere while the solution of the continuum problem itself is under investigation presently.

The paper is organized as follows. After necessary mathematical preliminaries and notations given in Section~\ref{mathprelim} we formulate the square lattice model in Section~\ref{squarelattice}. Here we apply the discrete Fourier transform to governing equations and present the transformed problem. In Section~\ref{ExactSolution} we present the exact solution for a strip of $N+1$ layers with fixed low boundary. Far-field behavior for lattice strip is discussed in Section~\ref{FarField} whereas the energy leakage flux ${\mathscr{D}}_{S}$ from the surface is analyzed in Section~\ref{Leakage}. Finally, we give some comments on the obtained results.

\section{Mathematical preliminaries}\label{mathprelim}
In what follows we use the following notation. 
Let ${\mathbb{Z}}$ denote the set of integers, ${{{\mathbb{Z}}^2}}$ denote ${\mathbb{Z}}\times{\mathbb{Z}}.$ Let ${\mathbb{Z}^+}$ denote the set of positive integers.
Let ${\mathbb{R}}$ denote the set of real numbers and ${\mathbb{C}}$ denote the set of complex numbers. The real part, $\Re {{z}}, $ of a complex number ${{z}}\in{\mathbb{C}}$ is denoted by ${{z}}_1\in{\mathbb{R}}$ and its imaginary part, $\Im {{z}}$, is denoted by ${{z}}_2\in{\mathbb{R}}$ (so that ${{z}}={{z}}_1+i{{z}}_2$). $|{{z}}|$ denotes the modulus and $\arg {{z}}$ denotes the argument (with branch cut along negative real axis) for ${{z}}\in{\mathbb{C}}$. If $f$ is a differentiable, real or complex, function then $f'$ denotes the derivative of $f$ with respect to its argument, and $f''$ denotes the second derivative. The discrete Fourier transform \cite{jury, Silbermann} ${\su}_{{\mathtt{y}}}^F: {\mathbb{C}}\to{\mathbb{C}}$ of $\{{\su}_{{\mathtt{x}}, {\mathtt{y}}}\}_{{\mathtt{x}}\in{\mathbb{Z}}}$ (along the ${\mathtt{x}}$ axis) is defined by
\begin{equation}\begin{split}
{\su}_{{\mathtt{y}}}^F={\su}_{{\mathtt{y}};+}+{\su}_{{\mathtt{y}};-}, \text{ where }
{\su}_{{\mathtt{y}};+}({{z}})=\sum\limits_{{\mathtt{x}}=0}^{+\infty} {\su}_{{\mathtt{x}}, {\mathtt{y}}}{{z}}^{-{\mathtt{x}}}, {\su}_{{\mathtt{y}};-}({{z}})=\sum\limits_{{\mathtt{x}}=-\infty}^{-1} {\su}_{{\mathtt{x}}, {\mathtt{y}}}{{z}}^{-{\mathtt{x}}}.\label{unpm}\end{split}\end{equation}
The symbol ${\mathbb{T}}$ denotes the unit circle (as a counterclockwise contour) in complex plane.
The letter ${{\mathcal{H}}}$ stands for the Heaviside function: ${{\mathcal{H}}}({\mathtt{x}})=0, {\mathtt{x}}<0$ and ${{\mathcal{H}}}({\mathtt{x}})=1, {\mathtt{x}}\ge0$.
The square root function, $\sqrt{\cdot}$, has the usual branch cut in the complex plane running from $-\infty$ to $0$. The symbol ${{z}}$ is exclusively used throughout as a complex variable for the discrete Fourier transform. The notation for relevant physical entities is described in the main text.

For scattering and Wiener-Hopf method we refer to 
\cite{Sommerfeld1,Sommerfeld2,Morse,Noble} whereas for 
convolution integrals and Fourier analysis to 
\cite{Paley, Wiener,Gakhov,KreinGoh}.

\section{Square lattice model: formulation for half-plane}\label{squarelattice}

\begin{figure}[!ht]
\center{\includegraphics[width=.9\textwidth]{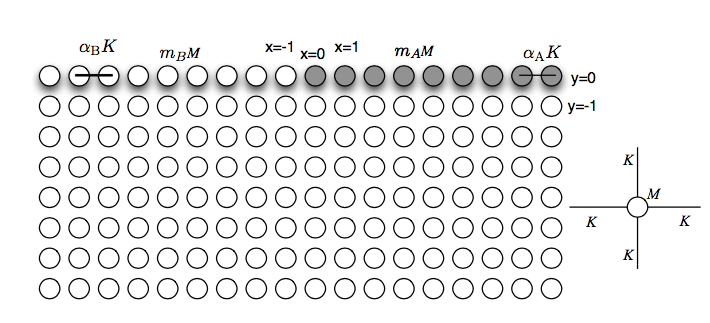}}
\caption[ ]{An illustration of the geometry of the lattice half space with structured surface interface on free boundary. A two dimensional projection is schematically illustrated from Fig. \ref{Fig1}.}
\label{Fig2}
\end{figure}

Let $\hat{\mathbf{i}}$ and $\hat{\mathbf{j}}$ form the standard (orthonormal) basis of ${\mathbb{R}}^2$.
Consider a two-dimensional semi-infinite (half-plane) square lattice
$\{\mathbf{x}\in{\mathbb{R}}^2: \mathbf{x}=m{a}\hat{\mathbf{i}}+n{a}\hat{\mathbf{j}}\text{ for some } m\in{\mathbb{Z}}, -n\in{\mathbb{Z}^+}\}.$
The lattice mostly consists of identical particles of mass $M$ connected to each other by linearly elastic bonds (springs) of stiffness $K$. In order to model surface tension we assume that the free surface ${\mathtt{y}}=0$ is constituted by particles with masses $m_s M$ and bonds with spring constant $\alpha_s K$, whereas $m_s$ and $\alpha_s$ are dimensionless parameters with $s={A}$ ahead of the surface coating junction and $s={B}$ behind. The anti-plane displacement of a particle, indexed by its lattice
coordinates ${\mathtt{x}}\in {\mathbb{Z}}$, ${\mathtt{y}}\in {\mathbb{Z}}$, is denoted by $\su_{{\mathtt{x}},{\mathtt{y}}}$.  
The motion equation for square lattice is given by
\begin{equation}\begin{split}\label{eqmotion}
 M \ddot{\su}_{{\mathtt{x}},{\mathtt{y}}}=K\left({\su}_{{\mathtt{x}}+1,{\mathtt{y}}}+{\su}_{{\mathtt{x}}-1,{\mathtt{y}}}+{\su}_{{\mathtt{x}},{\mathtt{y}}+1}+{\su}_{{\mathtt{x}},{\mathtt{y}}-1}-4{\su}_{{\mathtt{x}},{\mathtt{y}}}\right)
\end{split}\end{equation}
for ${\mathtt{x}}\in {{\mathbb{Z}^+}}$, ${\mathtt{y}}<0$, ${\mathtt{y}}\in {{\mathbb{Z}}}$, see, e.g. \cite{sharma2017linear}. On the free surface that is for $0<{\mathtt{x}}\in {{\mathbb{Z}}}$, ${\mathtt{y}}=0$ we have
\begin{equation}\begin{split}\label{eqmotionBCa}
 m_{{A}}M \ddot{\su}_{{\mathtt{x}},{\mathtt{y}}}=\alpha_{{A}} K\left({\su}_{{\mathtt{x}}+1,{\mathtt{y}}}+{\su}_{{\mathtt{x}}-1,{\mathtt{y}}}-2{\su}_{{\mathtt{x}},{\mathtt{y}}}\right)+K\left({\su}_{{\mathtt{x}},{\mathtt{y}}-1}-{\su}_{{\mathtt{x}},{\mathtt{y}}}\right),
\end{split}\end{equation}
for ${\mathtt{x}}\in {{\mathbb{Z}^-}}$, ${\mathtt{y}}=0$ we have
\begin{equation}\begin{split}\label{eqmotionBCb}
 m_{{B}}M \ddot{\su}_{{\mathtt{x}},{\mathtt{y}}}=\alpha_{{B}} K\left({\su}_{{\mathtt{x}}+1,{\mathtt{y}}}+{\su}_{{\mathtt{x}}-1,{\mathtt{y}}}-2{\su}_{{\mathtt{x}},{\mathtt{y}}}\right)+K\left({\su}_{{\mathtt{x}},{\mathtt{y}}-1}-{\su}_{{\mathtt{x}},{\mathtt{y}}}\right).
\end{split}\end{equation}
for ${\mathtt{x}}=0$, ${\mathtt{y}}=0$ we have
\begin{equation}\begin{split}\label{eqmotionBC0}
 m_{{A}}M \ddot{\su}_{{\mathtt{x}},{\mathtt{y}}}=\alpha_{{A}} K\left({\su}_{{\mathtt{x}}+1,{\mathtt{y}}}+{\su}_{{\mathtt{x}}-1,{\mathtt{y}}}-2{\su}_{{\mathtt{x}},{\mathtt{y}}}\right)+K\left({\su}_{{\mathtt{x}},{\mathtt{y}}-1}-{\su}_{{\mathtt{x}},{\mathtt{y}}}\right).
\end{split}\end{equation}

Motivated by a continuum context \citep{sharma2015diffraction}, let us re-write the symbols so that
\begin{equation}\begin{split}\label{defMKcTcS}
M=\rho {a}^3,\quad K=\mu {a},
\end{split}\end{equation}
where $\rho$ and $\mu$ are the mass density and shear modulus in the bulk.
The velocity of anti-plane shear waves in the continuum model is given by \cite{achenbach2012wave}
\begin{equation}\begin{split}
c=\sqrt{\mu/\rho}.
\end{split}\end{equation}
Thus, each particle of
the lattice half-plane possesses unit mass, and, interacts with {\em atmost} its four nearest neighbours through linearly elastic identical (massless) bonds with a spring constant $c^2/{a}^2$. The changes in lattice spacing near the boundary of the square lattice are ignored in this paper \citep{Maradudin,Wallis,Wallissurf} except for the change in mass and change in "horizontal" bond stiffness at the free surface.

Due to nature of the problem, the (out-of-plane) displacement of a particle, located at the site indexed by its {\em lattice coordinates} $({\mathtt{x}}, {\mathtt{y}})\in{{{\mathbb{Z}}^2}}$
 and denoted by ${\su}_{{\mathtt{x}}, {\mathtt{y}}}$, is complex valued. The notational devices employed in this paper find origin in the analysis of discrete Sommerfeld problems on the infinite square lattice \citep{sharma2015diffraction, sharma2015diffraction2,BlsNearCrack,BlsNearRigid}.

The equation of motion of the portion of lattice half-plane away from its boundary is
\begin{equation}\begin{split}
({a}/{c})^2\ddot{\su}_{{\mathtt{x}}, {\mathtt{y}}}=&{\triangle}{\su}_{{\mathtt{x}}, {\mathtt{y}}},
\text{where }
{\triangle}{\su}_{{\mathtt{x}}, {\mathtt{y}}}\equiv{\su}_{{\mathtt{x}}+1, {\mathtt{y}}}+{\su}_{{\mathtt{x}}-1, {\mathtt{y}}}+{\su}_{{\mathtt{x}}, {\mathtt{y}}+1}+{\su}_{{\mathtt{x}}, {\mathtt{y}}-1}-4{\su}_{{\mathtt{x}}, {\mathtt{y}}},
\label{newtoneq_sq}
\end{split}\end{equation}
while on each semi-infinite piece of the half-plane boundary modulo $\{0,0\}$ and the site at ${\mathtt{x}}=0$,
it is assumed that
\begin{subequations}
\begingroup
\addtolength{\jot}{0em}
\begin{align}
m_{{A}}({a}/{c})^2\ddot{\su}_{{{\mathtt{x}}}, {{\mathtt{y}}}}&=\alpha_{{A}}({\su}_{{{\mathtt{x}}}+1, {{\mathtt{y}}}}+{\su}_{{{\mathtt{x}}}-1, {{\mathtt{y}}}}-2{\su}_{{{\mathtt{x}}}, {{\mathtt{y}}}})+{\su}_{{{\mathtt{x}}}, {{\mathtt{y}}}-1}-{\su}_{{{\mathtt{x}}}, {{\mathtt{y}}}}&&\text{($s={A}$)}\label{bc_sq_A}\\
m_{{B}}({a}/{c})^2\ddot{\su}_{{{\mathtt{x}}}, {{\mathtt{y}}}}=&\alpha_{{B}}({\su}_{{{\mathtt{x}}}+1, {{\mathtt{y}}}}+{\su}_{{{\mathtt{x}}}-1, {{\mathtt{y}}}}-2{\su}_{{{\mathtt{x}}}, {{\mathtt{y}}}})+{\su}_{{{\mathtt{x}}}, {{\mathtt{y}}}-1}-{\su}_{{{\mathtt{x}}}, {{\mathtt{y}}}}&&\text{($s={B}$)}.\label{bc_sq_B}\\
m_{{A}}({a}/{c})^2\ddot{\su}_{{{\mathtt{x}}}, {{\mathtt{y}}}}&=\alpha_{{B}}({\su}_{{{\mathtt{x}}}-1, {{\mathtt{y}}}}-{\su}_{{{\mathtt{x}}}, {{\mathtt{y}}}})+\alpha_{{A}}({\su}_{{{\mathtt{x}}}+1, {{\mathtt{y}}}}-{\su}_{{{\mathtt{x}}}, {{\mathtt{y}}}})+{\su}_{{{\mathtt{x}}}, {{\mathtt{y}}}-1}-{\su}_{{{\mathtt{x}}}, {{\mathtt{y}}}}&&\text{(${\mathtt{x}}=0$)}\label{bc_sq_0}.
\end{align}
\endgroup
\end{subequations}

Let us consider the incident surface wave
\begin{equation}\begin{split}
{\su}^{\inc}_{{\mathtt{x}},{\mathtt{y}}}={{\mathrm{A}}}\exp({-i{\upxi}_{\inc} {\mathtt{x}}-i\omega t})\exp({\eta_{\inc} {\mathtt{y}}}),
\label{uincwave}
\end{split}\end{equation}
where ${\upxi}_{\inc}$ is the discrete wave number from the right side such that ${\mathtt{V}_g}({\upxi}_{\inc})<0$, ${\upxi}_{\inc}\in(0,\pi)$, and $\eta_{\inc}$ is assumed to be positive. A detailed description of the surface waves can be found in \cite{Victor_Bls_surf1}.
Let
\begin{equation}\begin{split}\label{deffreq}
{\upomega}=\omega{a}/{c}.
\end{split}\end{equation}
It is found that ${\upomega}={\upomega}_{{A}}({\upxi}_{\inc})$ and $\eta_{\inc}=\eta_{{A}}({\upxi}_{\inc})$ satisfy the two equations
\begin{align}\label{eq1}
 &-{\upomega}^2=\left(2\cos{\upxi}_{\inc}+2\cosh\eta_{\inc}-4\right),
 \\\label{eq2}&-m_{{A}}{\upomega}^2=\alpha_{{A}} \left(2\cos{\upxi}_{\inc}-2\right)+ (\exp(-\eta_{\inc})-1).
\end{align}
Assuming the steady state we look for the constant frequency solutions of \eqref{newtoneq_sq} and \eqref{bc_sq_A}--\eqref{bc_sq_0}. Let $\su_{{\mathtt{x}}, {\mathtt{y}}}$ denote the scattered wavefield, i.e., $\su_{{\mathtt{x}}, {\mathtt{y}}}+\su^{\inc}_{{\mathtt{x}}, {\mathtt{y}}}$ represents the total wavefield at the site $({\mathtt{x}}, {\mathtt{y}})$. Using the analysis of \cite{sharma2015diffraction} and \cite{sharma2017scattering} it is easy to see that the Fourier transform of $\su_{{\mathtt{x}}, {\mathtt{y}}}$ in the half space can be written as
\begin{equation}\begin{split}\label{halfspace}
\su^F_{{\mathtt{y}}}=\su^F_0{\lambda}^{-{\mathtt{y}}}, {\mathtt{y}}\le0,
\end{split}\end{equation}
with
\begin{equation}
\lambda({z})= \frac{{\mathtt{r}}({z})-{\mathtt{h}}({z})}{{\mathtt{r}}({z})+{\mathtt{h}}({z})},
\label{lambda}
\end{equation}
where
\citep{slepyan2012models}
\begin{equation}
{\mathtt{h}}({z})=\sqrt{{\mathtt{Q}}({z})-2},\quad {\mathtt{r}}({z})=\sqrt{{\mathtt{Q}}({z})+2},\quad {\mathtt{Q}}({z})=4-{z}-{z}^{-1}-{\upomega}^2.
\label{defQ}
\end{equation}
The Fourier transform as well as other functions stated above (according to the summation \eqref{unpm}) are analytic in a suitable annulus ${\mathfrak{A}}.$
It can be shown that the annulus ${\mathfrak{A}}$ contains the unit circle ${\mathbb{T}}$, i.e., ${\mathbb{T}}=\{|{z}|=1| {z}\in{\mathbb{C}}\}$; we omit the details.
For convenience, we introduce the notation that $\su^F_0=\su^F$. Let
\begin{equation}\begin{split}\label{defFT}
\su^F=\su_++\su_-,
\end{split}\end{equation}
where $\su_+$ is analytic on the annulus ${\mathfrak{A}}$ and outside it while $\su_-$ is analytic on the annulus ${\mathfrak{A}}$ and inside it.
According to \eqref{bc_sq_A}--\eqref{bc_sq_0}, since $\su^{\inc}_{{\mathtt{x}}, {\mathtt{y}}}$ \eqref{uincwave} satisfies the equation of motion in the portion with boundary with index ${s}={A}$, for ${\mathtt{y}}=0$,
\begin{equation}\begin{split}\label{BCeq}
&\alpha_{{A}}({\su}_{{{\mathtt{x}}}+1, {{\mathtt{y}}}}-{\su}_{{{\mathtt{x}}}, {{\mathtt{y}}}}){{\mathcal{H}}}({\mathtt{x}})
+\alpha_{{A}}({\su}_{{{\mathtt{x}}}-1, {{\mathtt{y}}}}-{\su}_{{{\mathtt{x}}}, {{\mathtt{y}}}}){{\mathcal{H}}}({\mathtt{x}}-1)\\
&+\alpha_{{B}}({\su}_{{{\mathtt{x}}}+1, {{\mathtt{y}}}}-{\su}_{{{\mathtt{x}}}, {{\mathtt{y}}}}){{\mathcal{H}}}(-{\mathtt{x}}-1)
+\alpha_{{B}}({\su}_{{{\mathtt{x}}}-1, {{\mathtt{y}}}}-{\su}_{{{\mathtt{x}}}, {{\mathtt{y}}}}){{\mathcal{H}}}(-{\mathtt{x}})\\
&+{\su}_{{{\mathtt{x}}}, {{\mathtt{y}}}-1}-{\su}_{{{\mathtt{x}}}, {{\mathtt{y}}}}+m_{{A}}{\upomega}^2{\su}_{{{\mathtt{x}}}, {{\mathtt{y}}}}{{\mathcal{H}}}({\mathtt{x}})+m_{{B}}{\upomega}^2{\su}_{{{\mathtt{x}}}, {{\mathtt{y}}}}{{\mathcal{H}}}(-{\mathtt{x}}-1)=-f^{\inc}_{{\mathtt{x}}},
\end{split}\end{equation}
where (with ${\mathtt{y}}=0$)
\begin{equation}\begin{split}\label{deffinc}
f^{\inc}_{{\mathtt{x}}}
=(\alpha_{{B}}-\alpha_{{A}})({\su}^{\inc}_{{{\mathtt{x}}}+1, {{\mathtt{y}}}}-{\su}^{\inc}_{{{\mathtt{x}}}, {{\mathtt{y}}}}){{\mathcal{H}}}(-{\mathtt{x}}-1)\\
+(\alpha_{{B}}-\alpha_{{A}})({\su}^{\inc}_{{{\mathtt{x}}}-1, {{\mathtt{y}}}}-{\su}^{\inc}_{{{\mathtt{x}}}, {{\mathtt{y}}}}){{\mathcal{H}}}(-{\mathtt{x}})\\
+(m_{{B}}-m_{{A}}){\upomega}^2{\su}^{\inc}_{{{\mathtt{x}}}, {{\mathtt{y}}}}{{\mathcal{H}}}(-{\mathtt{x}}-1).
\end{split}\end{equation}
The Fourier transform of above equation leads to
\begin{equation}\begin{split}\label{BCeqF}
\alpha_{{A}}({z}{\su}_+-{z}\su_{0, 0}-{\su}_+)+\alpha_{{A}}({z}^{-1}{\su}_++\su_{0, 0}-{\su}_+)\\
+\alpha_{{B}}({z}{\su}_-+{z}\su_{0, 0}-{\su}_-)+\alpha_{{B}}({z}^{-1}{\su}_--\su_{0, 0}-{\su}_-)\\
+({\su}_++{\su}_-)({\lambda}-1)+m_{{A}}{\upomega}^2{\su}_++m_{{B}}{\upomega}^2{\su}_-=-\sum\limits_{{\mathtt{x}}\in{\mathbb{Z}}}f^{\inc}_{{\mathtt{x}}}{z}^{-{\mathtt{x}}},
\end{split}\end{equation}
where the right hand side can be found to be
\begin{equation}\begin{split}
-(\alpha_{{B}}-\alpha_{{A}})({z}{\su}^{\inc}_-+{z}\su^{\inc}_{0, 0}-{\su}^{\inc}_-+{z}^{-1}{\su}^{\inc}_--\su^{\inc}_{0, 0}-{\su}^{\inc}_-)-(m_{{B}}-m_{{A}}){\upomega}^2{\su}^{\inc}_-\\=
-(\alpha_{{B}}-\alpha_{{A}})(({z}+{z}^{-1}-2){\su}^{\inc}_-+({z}-1)\su^{\inc}_{0, 0})-(m_{{B}}-m_{{A}}){\upomega}^2{\su}^{\inc}_-
\label{discWHkerpre}
\end{split}\end{equation}
where (using \eqref{uincwave})
\begin{equation}\begin{split}
\su^{\inc}_-={\mathrm{A}}\delta_{D-}({z}{z}_{\inc}^{-1}),\quad {z}_{\inc}=e^{-i{\upxi}_{\inc}}.
\label{uincF}
\end{split}\end{equation}
Due to \eqref{uincwave},
\begin{equation}\begin{split}\su^{\inc}_-({z})=\sum_{{\mathtt{x}}\in\mathbb{Z}^-}{\su}^{\inc}_{{\mathtt{x}},{\mathtt{y}}}{z}^{-{\mathtt{x}}}={\mathrm{A}}\delta_{D-}({z}{z}_{\inc}^{-1})=-{\mathrm{A}}\frac{{z}}{{z}-{z}_{\inc}}, \quad |{z}{z}_{\inc}^{-1}|<1.\end{split}\end{equation}
Note that $|{z}_{\inc}|=|e^{-i{\upxi}_{\inc}}|=|e^{\Im{\upxi}_{\inc}}|>1$ assuming a vanishing amount of damping so that ${\upomega}={\upomega}_1+i{\upomega}_2, 0<{\upomega}_2\ll1$.
With
${s}={A}, {B}$,
let
\begin{equation}\begin{split}
{\mathfrak{K}}_{s}={\lambda}+F_s,
\label{discWHker}
\end{split}\end{equation}
where
\begin{equation}\begin{split}
F_s({z})=m_{s}{\upomega}^2-1+\alpha_{s}({z}+{z}^{-1}-2).
\label{discWHF1}
\end{split}\end{equation}
$F_s$ is a function of the same form as ${\mathtt{Q}}$ \eqref{defQ}.
Collecting the terms accompanying $\su_{\pm}$ in \eqref{discWHkerpre},
we get
\begin{equation}\begin{split}
{\mathfrak{K}}_{{A}}{\su}_++{\mathfrak{K}}_{{B}}{\su}_-=
-(\alpha_{{B}}-\alpha_{{A}})(({z}+{z}^{-1}-2){\su}^{\inc}_-+({z}-1)\su^{\inc}_{0, 0})\\
-(m_{{B}}-m_{{A}}){\upomega}^2{\su}^{\inc}_--(\alpha_{{B}}-\alpha_{{A}})({z}-1)\su_{0, 0},
\label{discWHkerpre1}
\end{split}\end{equation}
Note that ${\mathfrak{K}}_{{A}}-{\mathfrak{K}}_{{B}}=(\alpha_{{A}}-\alpha_{{B}})({z}+{z}^{-1}-2)+(m_{{A}}-m_{{B}}){\upomega}^2$ and ${\mathfrak{K}}_{{A}}({z}_{\inc})=0$.
Hence, \eqref{discWHkerpre1} can be re-written as
\begin{equation}\begin{split}
{\mathfrak{K}}_{{A}}{\su}_++{\mathfrak{K}}_{{B}}{\su}_-=
({\mathfrak{K}}_{{A}}-{\mathfrak{K}}_{{B}}){\su}^{\inc}_--(\alpha_{{A}}-\alpha_{{B}})(1-{z})\su^{\totwave}_{0, 0}.
\label{discWH}
\end{split}\end{equation}
The Wiener-Hopf (WH) kernel in \eqref{discWH} is ${\mathfrak{K}}_{{B}}/{\mathfrak{K}}_{{A}}$ which needs to be factorized multiplicatively.
Thus, the half space problem involves a multiplicative factorization problem which requires further analysis and computations to a reasonable extent, a closed form solution is formidable task.
It is indeed serendipity that a variant of the problem for the lattice half-plane does admit a closed form solution, this can be construed as a kind of approximate factorization for the relevant WH problem \eqref{discWH} as well. This is discussed in the next section.

\section{Exact solution for lattice strip}\label{ExactSolution}

\begin{figure}[htb!]
\center{\includegraphics[width=.9\textwidth]{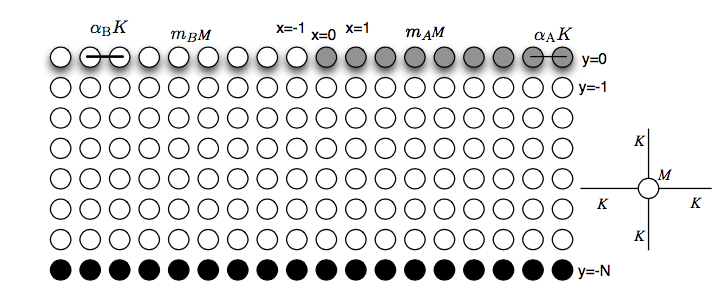}}
\caption[ ]{An illustration of the geometry of the lattice strip with structured surface interface on free boundary.}
\label{Fig3}
\end{figure}

\begin{figure}[htb!]
\center{\includegraphics[width=\textwidth]{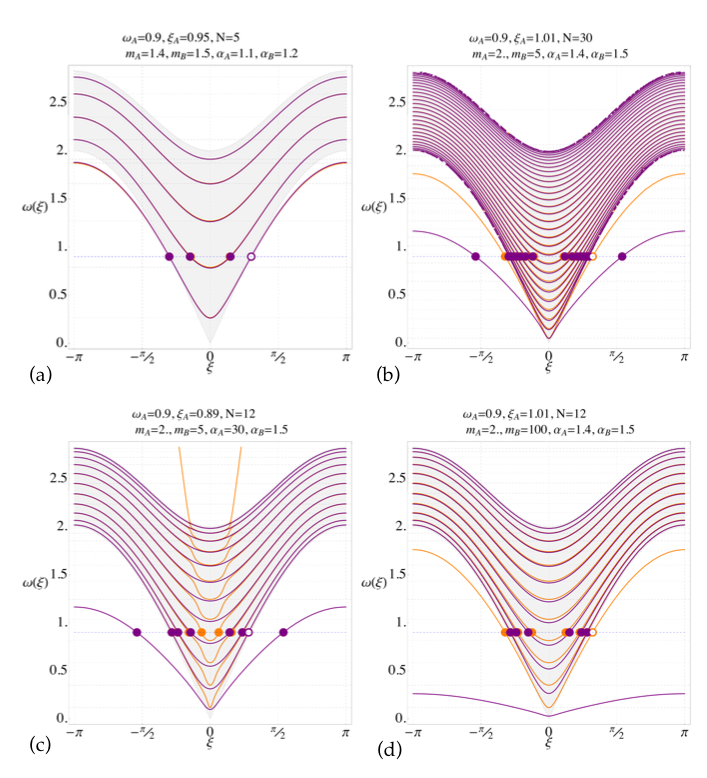}}
\caption[ ]{Illustration of the dispersion relation \eqref{surfacewavedispersion} for the lattice strip with structured surface interface on free boundary. ${A}$ stands for the portion ahead of the interface while ${B}$ stands for the one behind.}
\label{Fig4}
\end{figure}

Suppose that the lower surface of the square lattice strip is fixed, i.e.,
\begin{equation}\begin{split}\label{latticestripBC}
\su^{\totwave}_{{\mathtt{x}}, {\mathtt{y}}}=0\text{ at }{\mathtt{y}}=-\mathtt{N}\text{ for all }{\mathtt{x}}\in{\mathbb{Z}}.
\end{split}\end{equation}
Clearly the appropriate incident wave also satisfies $\su^{\inc}_{{\mathtt{x}}, {\mathtt{y}}}=0$ at ${\mathtt{y}}=-\mathtt{N}$ while it satisfies the free surface condition for ${s}={A}$. The dispersion relation for such a wave can be also easily found (see Fig. \ref{Fig4} for some choices of parameters). The relevant details are provided in \ref{app_surface_strip}. It is easy to see that the incident wave form when evaluated at ${\mathtt{y}}=0$ retains the same form so that \eqref{uincF} persists; for specific purpose, it is given by (using \eqref{surfwavegen})
\begin{equation}\begin{split}
{\su}^{\inc}_{{\mathtt{x}},{\mathtt{y}}}={{\mathrm{A}}}{\hat{a}}_{{A};{\mathtt{y}}}({z}_{\inc})\exp({-i{\upxi}_{\inc} {\mathtt{x}}-i\omega t}),
\label{uincwaveS}
\end{split}\end{equation}
where ${\upxi}_{\inc}$ is the discrete wave number from the right side such that ${\mathtt{V}_g}({\upxi}_{\inc})<0$, ${\upxi}_{\inc}\in(0,\pi)$.

In place of the general solution \eqref{halfspace} in the lattice half space, the expression is given by
\begin{equation}\begin{split}\label{latticestrip}
\su^F_{{\mathtt{y}}}=\su^F{\mathfrak{V}},\quad {\mathtt{y}}\le0,
\end{split}\end{equation}
where
\begin{equation}\begin{split}\label{latticestrip2}
{\mathfrak{V}}_{{\mathtt{y}}}=\frac{{\lambda}^{-{\mathtt{y}}-\mathtt{N}}-{\lambda}^{{\mathtt{y}}+\mathtt{N}}}{{\lambda}^{-\mathtt{N}}-{\lambda}^{\mathtt{N}}}.
\end{split}\end{equation}
In term of the Chebyshev polynomials \citep{Mason} and using the recent analysis provided by \citep{sharma2017linear} and \cite{Bifurcated}, note that
\begin{equation}\begin{split}
\su^F_{-1}=\su^F\Lambda, \quad\Lambda={\mathfrak{V}}_1=({{\lambda}^{1-\mathtt{N}}-{\lambda}^{-1+\mathtt{N}}})/({{\lambda}^{-\mathtt{N}}-{\lambda}^{\mathtt{N}}})={\mathtt{U}}_{\mathtt{N}-2}/{\mathtt{U}}_{\mathtt{N}-1}.
\end{split}\end{equation}
Naturally,
the Wiener-Hopf equation
\eqref{discWH} is obtained again with
a change that \eqref{discWHker} is replaced by
\begin{equation}\begin{split}
{\mathfrak{K}}_{s}=\Lambda+F_{s},
\label{discWHker2}
\end{split}\end{equation}
while $F_s$ continues to be given by \eqref{discWHF1}. Unlike the case of \eqref{discWHker}, the expressions of ${\mathfrak{K}}_{{A}}$ and ${\mathfrak{K}}_{{B}}$ are found to be polynomials involving ${z}$ and ${z}^{-1}$ so that the symbolic framework developed by \cite{Bifurcated} applies.
The Wiener-Hopf equation
\eqref{discWH} becomes
\begin{equation}\begin{split}
{\su}_++\frac{{\mathfrak{K}}_{{B}}}{{\mathfrak{K}}_{{A}}}{\su}_-=
(1-\frac{{\mathfrak{K}}_{{B}}}{{\mathfrak{K}}_{{A}}}){\su}^{\inc}_--(\alpha_{{A}}-\alpha_{{B}})\su^{\totwave}_{0, 0}\frac{(1-{z})}{{\mathfrak{K}}_{{A}}}.
\label{discWHstrip}
\end{split}\end{equation}
Let the Wiener-Hopf kernel be denoted by
\begin{equation}\begin{split}
{\mathfrak{L}}=\frac{{\mathfrak{K}}_{{B}}}{{\mathfrak{K}}_{{A}}}=\frac{\Lambda+F_{{B}}}{\Lambda+F_{{A}}}=\frac{{\mathtt{U}}_{\mathtt{N}-2}+F_{{B}}{\mathtt{U}}_{\mathtt{N}-1}}{{\mathtt{U}}_{\mathtt{N}-2}+F_{{A}}{\mathtt{U}}_{\mathtt{N}-1}}=\frac{{\mathcal{B}}}{{\mathcal{A}}}.
\label{discWHker1}
\end{split}\end{equation}
Then \eqref{discWHstrip} can be re-written as
\begin{equation}\begin{split}
{\su}_++{\mathfrak{L}}{\su}_-=
(1-{\mathfrak{L}}){\su}^{\inc}_--(\alpha_{{A}}-\alpha_{{B}})\su^{\totwave}_{0, 0}\frac{(1-{z})}{{\mathfrak{K}}_{{A}}}.
\label{discWHker12}
\end{split}\end{equation}
Here ${\mathfrak{K}}_{{A}}({z}^{-1}_{\inc})=0$, in fact, ${\mathfrak{K}}_{{A}+}({z}^{-1}_{\inc})=0$ as $|{z}_{\inc}|>1$ so that ${\mathfrak{L}}^{-1}_+({z}^{-1}_{\inc})=0$ (${\mathcal{A}}_+({z}^{-1}_{\inc})=0$).
It is easy to see that the multiplicative factorization of ${\mathcal{B}}$ and ${\mathcal{A}}$, which are polynomials in ${z}$ and ${z}^{-1}$, can be carried out as spelled in \cite{Bifurcated}, i.e.,
\begin{equation}\begin{split}
{\mathfrak{L}}={\mathfrak{L}}_+{\mathfrak{L}}_-=\frac{{\mathcal{B}}_+{\mathcal{B}}_-}{{\mathcal{A}}_+{\mathcal{A}}_-}.
\label{discWHker3}
\end{split}\end{equation}
Then \eqref{discWHker12} can be re-written as (note that ${\mathfrak{K}}_{{A}}={\mathcal{A}}/{\mathtt{U}}_{\mathtt{N}-1}={\mathcal{A}}_+{\mathcal{A}}_-/{\mathtt{U}}_{\mathtt{N}-1}$)
\begin{equation}\begin{split}
{\mathfrak{L}}_{+}^{-1}{\su}_++{\mathfrak{L}}_-{\su}_-={\mathfrak{C}}={\mathfrak{C}}_++{\mathfrak{C}}_-,
\label{discWHN2}
\end{split}\end{equation}
with
\begin{equation}\begin{split}
{\mathfrak{C}}({z})
&=({\mathfrak{L}}_{+}^{-1}-{\mathfrak{L}}_-){\su}^{\inc}_--(\alpha_{{A}}-\alpha_{{B}})\su^{\totwave}_{0, 0}\mathcal{F}_{+}({z})\mathcal{G}_{-}({z}),
\label{defCeq1}
\end{split}\end{equation}
where
\begin{equation}\begin{split}
F_{\mathtt{N}-1;+}&={\mathtt{U}}_{\mathtt{N}-1;+},\quad
G_{\mathtt{N};-}=(1-{z}){\mathtt{U}}_{\mathtt{N}-1;-},\\
\mathcal{F}_{+}({z})&=\frac{F_{\mathtt{N}-1;+}({z})}{{\mathcal{B}}_{\mathtt{N};+}({z})}=\frac{F_{\mathtt{N}-1;+}({z})}{{\mathcal{B}}_{+}({z})},\\
\mathcal{G}_{-}&=\frac{G_{\mathtt{N};-}}{{\mathcal{A}}_{\mathtt{N};-}}=\frac{G_{\mathtt{N};-}}{{\mathcal{A}}_{-}}.
\end{split}\end{equation}
The rational function $\mathcal{F}_{+}({z})$ can be expanded into partial fractions \citep{Ablowitz} using the Cauchy residue theorem. The denominator of $\mathcal{F}_{+}({z})$ is ${\mathcal{B}}_{+}({z})$ which can be written as product of terms of type $\sqrt{{z}_i}^{-1}(1-{z}_i/{z})\equiv\sqrt{{z}_i}^{-1}{z}^{-1}({z}-{z}_i)$ ($i$ ranging over the number of zeros of $\mathcal{B}$ inside the unit circle $\mathbb{T}$ in the complex plane).
Assume that ${\upomega}$ is such that ${\mathcal{B}}_+$ has only simple poles (a generic case, indeed). Note that
\begin{equation}\begin{split}
\lim_{{{z}}\to0}\mathcal{F}_{+}({z})=\lim_{{{z}}\to0}\dfrac{F_{\mathtt{N}-1;+}({z})}{{\mathcal{B}}_{+}({z})}=\lim_{{{z}}\to0}\dfrac{{z}^\mathtt{N}{\mathtt{U}}_{\mathtt{N}-1;+}}{{z}^\mathtt{N}{\mathcal{B}}_{+}({z})}\propto{z}\to0.
\end{split}\end{equation}
Above observation allows the expansion
\begin{equation}\begin{split}
\frac{1}{{z}}\mathcal{F}_{+}({z})=
\sum\limits_{\widetilde{{z}}|{\mathcal{B}}_+(\widetilde{{z}})=0}\frac{1}{\widetilde{{z}}}\frac{\text{Res }|_{\widetilde{{z}}}\mathcal{F}_{+}({z})}{{z}-\widetilde{{z}}}=\sum\limits_{\widetilde{{z}}|{\mathcal{B}}_+(\widetilde{{z}})=0}{\mathcal{P}}({\widetilde{{z}}})\frac{1}{{z}-\widetilde{{z}}},
\end{split}\end{equation}
\text{where }
\begin{equation}\begin{split}
{\mathcal{P}}({\widetilde{{z}}})=\frac{1}{\widetilde{{z}}}\text{Res }|_{\widetilde{{z}}}\mathcal{F}_{+}({z})=\frac{1}{\widetilde{{z}}}\frac{F_{\mathtt{N}-1;+}(\widetilde{{z}})}{{\mathcal{B}}'_{+}(\widetilde{{z}})}.
\label{defAzt}
\end{split}\end{equation}
Thus, \eqref{defCeq1} leads to
\begin{equation}\begin{split}
{\mathfrak{C}}({z})
&=({\mathfrak{L}}_{+}^{-1}({z})-{\mathfrak{L}}_-({z})){\su}^{\inc}_-({z})-(\alpha_{{A}}-\alpha_{{B}})\su^{\totwave}_{0, 0}\sum\limits_{\widetilde{{z}}|{\mathcal{B}}_+(\widetilde{{z}})=0}{\mathcal{P}}({\widetilde{{z}}})\frac{{z}}{{z}-\widetilde{{z}}}\mathcal{G}_{-}({z}).
\end{split}\end{equation}
Due to the assumed form of the incident wave \eqref{uincwaveS},
\begin{equation}\begin{split}\su^{\inc}_-({z})={{\mathrm{A}}{\hat{a}}_{{A};0}({z}_{\inc})}\delta_{D-}({z}{z}_{\inc}^{-1})
=-{{\mathrm{A}}{\hat{a}}_{{A};0}({z}_{\inc})}\frac{{z}}{{z}-{z}_{\inc}}.\end{split}\end{equation}
Indeed, the manipulations outlined so far lead to the additive factorization of ${\mathfrak{C}}$ such that ${\mathfrak{C}}={\mathfrak{C}}_++{\mathfrak{C}}_-$ where
\begin{equation}\begin{split}
{\mathfrak{C}}_-({z})&={{\mathrm{A}}{\hat{a}}_{{A};0}({z}_{\inc})}({\mathfrak{L}}_-({z})-{\mathfrak{L}}_+^{-1}({z}_{\inc}))\frac{{z}}{{z}-{z}_{\inc}}\\
&-(\alpha_{{A}}-\alpha_{{B}})\su^{\totwave}_{0, 0}\sum\limits_{\widetilde{{z}}|{\mathcal{B}}_+(\widetilde{{z}})=0}{\mathcal{P}}({\widetilde{{z}}})\frac{{z}}{{z}-\widetilde{{z}}}(\mathcal{G}_{-}({z})-\mathcal{G}_{-}(\widetilde{{z}})),
\end{split}\end{equation}
and
\begin{equation}\begin{split}
{\mathfrak{C}}_+({z})&={{\mathrm{A}}{\hat{a}}_{{A};0}({z}_{\inc})}(-{\mathfrak{L}}_+^{-1}({z})+{\mathfrak{L}}_+^{-1}({z}_{\inc}))\frac{1}{1-{z}_{\inc}{z}^{-1}}\\
&-(\alpha_{{A}}-\alpha_{{B}})\su^{\totwave}_{0, 0}\sum\limits_{\widetilde{{z}}|{\mathcal{B}}_+(\widetilde{{z}})=0}{\mathcal{P}}({\widetilde{{z}}})\frac{\mathcal{G}_{-}(\widetilde{{z}})}{1-\widetilde{{z}}{z}^{-1}}.
\end{split}\end{equation}
${\mathfrak{C}}_+$ is analytic on the annulus ${\mathfrak{A}}$ and outside it while ${\mathfrak{C}}_+$ is analytic on the annulus ${\mathfrak{A}}$ and inside it.
As $|\widetilde{{z}}|<1,$ note that ${\mathfrak{L}}_-({z}){\su}_-({z})\to0, {\mathfrak{C}}_-({z})\to{\mathfrak{C}}_-(0)=0,$ a constant as ${z}\to0$ and also note that ${\mathfrak{L}}_{+}^{-1}({z}){\su}_+({z})\to$ a constant, ${\mathfrak{C}}_+({z})\to{\mathfrak{C}}_+(\infty)$ a constant as ${z}\to\infty$. Hence,  \eqref{discWHN2} can be split into the form
\begin{equation}\begin{split}
{\mathfrak{L}}_{+}^{-1}{\su}_+-{\mathfrak{C}}_+={\mathfrak{C}}_--{\mathfrak{L}}_-{\su}_-={\mathfrak{C}}_-(0)=0.
\label{expnCn0}
\end{split}\end{equation}
For the last equation in \eqref{expnCn0}, we use the Lioville's theorem \citep{Ablowitz,Noble}.
Finally,
\begin{equation}\begin{split}
{\su}_+={\mathfrak{L}}_{+}{\mathfrak{C}}_+, \quad {\su}_-={\mathfrak{L}}_-^{-1}{\mathfrak{C}}_-,
\label{discWHNsol}
\end{split}\end{equation}
and after adding both half-Fourier transforms
\begin{equation}\begin{split}
{\su}^F({z})&={\mathfrak{L}}_{+}({z}){\mathfrak{C}}_+({z})+{\mathfrak{L}}_-^{-1}({z}){\mathfrak{C}}_-({z})\\
&=-\frac{{{\mathrm{A}}{\hat{a}}_{{A};0}({z}_{\inc})}}{{\mathfrak{L}}_+({z}_{\inc})}\frac{1-{\mathfrak{L}}({z})}{{\mathfrak{L}}_-({z})}\frac{{z}}{{z}-{z}_{\inc}}\\
&-(\alpha_{{A}}-\alpha_{{B}})\su^{\totwave}_{0, 0}\sum\limits_{\widetilde{{z}}|{\mathcal{B}}_+(\widetilde{{z}})=0}{\mathcal{P}}({\widetilde{{z}}})\frac{{z}}{{z}-\widetilde{{z}}}(\frac{\mathcal{G}_{-}({{z}})}{{\mathfrak{L}}_-({z})}-\mathcal{G}_{-}(\widetilde{{z}})\frac{1-{\mathfrak{L}}({z})}{{\mathfrak{L}}_-({z})}).
\label{discWHNsolfull}
\end{split}\end{equation}

\begin{figure}[htb!]
\center{\includegraphics[width=.7\textwidth]{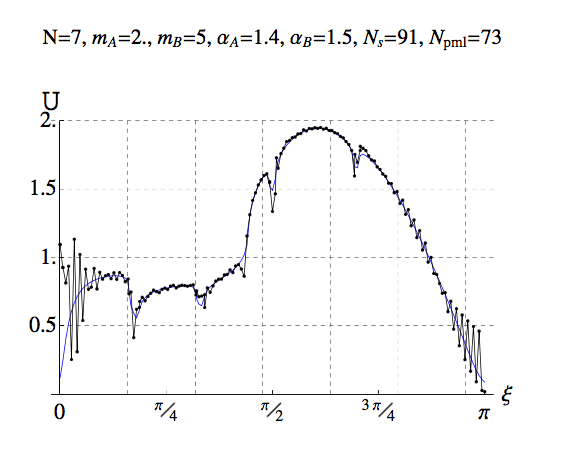}}
\caption[ ]{Illustration of the evaluation of $\mathcal{U}$ given by \eqref{ut00analytical}. The blue curve refers to the analytical expression \eqref{ut00analytical} while the black curve (with points) refers to the value obtained via a numerical solution of the scattering problem.}
\label{Fig5}
\end{figure}

\begin{figure}[htb!]
\center{\includegraphics[width=.9\textwidth]{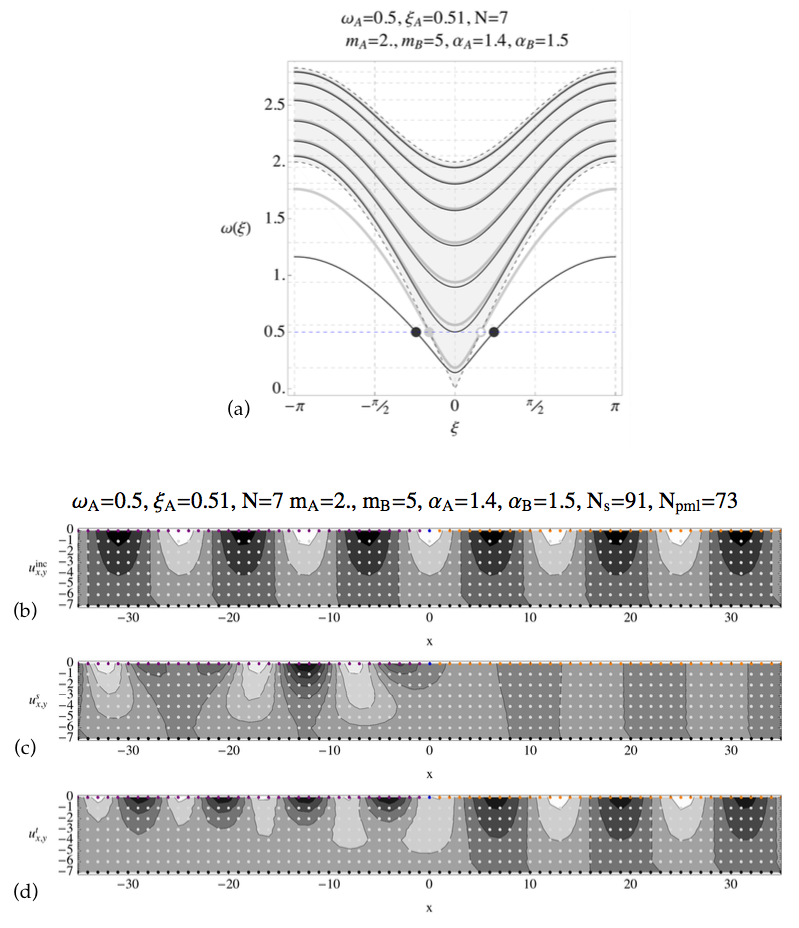}}
\caption[ ]{(a) Illustration of the dispersion relation for the lattice strip with surface structure on free boundary. In the dispersion curves, the color gray refers to right part ${s}={A}$ while the color black refers to left part ${s}={B}.$ ${\upomega}_{{A}}={\upomega}_{\inc}={\upomega}({\upxi}_{\inc})$. The gray shaded region is the pass band of the bulk lattice. The bottom three plots (b), (c), and (d) present the contourplots of the real part of the incident wave $\su^{\inc}_{{\mathtt{x}}, {\mathtt{y}}}$, scattered wave field $\su^{s}_{{\mathtt{x}}, {\mathtt{y}}}$ ($=\su_{{\mathtt{x}}, {\mathtt{y}}}$) and the total wave field $\su^{\totwave}$, respectively. The dots superpose the physical lattice strip with the contour plot (horizontal axis is $\mathtt{x}$ and vertical axis is $\mathtt{y}$.}
\label{Fig6}
\end{figure}

Note that $\su^{\totwave}_{0, 0}$, i.e., $\su_{0, 0}$, is still an unknown at this stage. Using the technique similar to that for the site facing the rigid constraint \cite{sharma2015diffraction2}, it is found that
\begin{equation}\begin{split}
\su^{\totwave}_{0, 0}={{{\mathrm{A}}{\hat{a}}_{{A};0}({z}_{\inc})}}\mathcal{U}({z}_{\inc}),
\end{split}\end{equation}
where $\mathcal{U}({z}_{\inc})$ is given by
\begin{equation}\begin{split}
\mathcal{U}({z}_{\inc})=\dfrac{1}{{\mathfrak{L}}_+({z}_{\inc})}\dfrac{\sum\limits_{{z}|{\mathcal{B}}_-({z})=0}\dfrac{{\mathcal{A}}_-({z})}{{\mathcal{B}}'_-({z})}\dfrac{1}{{z}-{z}_{\inc}}}{1-(\alpha_{{A}}-\alpha_{{B}})\sum\limits_{\widetilde{{z}}|{\mathcal{B}}_+(\widetilde{{z}})=0}\sum\limits_{{z}|{\mathcal{B}}_-({z})=0}\dfrac{{\mathcal{P}}({\widetilde{{z}}})}{{z}-\widetilde{{z}}}\dfrac{G_{\mathtt{N};-}({{z}})-\mathcal{G}_{-}(\widetilde{{z}}){\mathcal{A}}_-({z})}{{\mathcal{B}}'_-({z})}},
\label{ut00analytical}
\end{split}\end{equation}
which is derived in the \ref{app_u00}. The expression \eqref{ut00analytical} has been verified analytically by a numerical solution of the scattering problem, see Fig. \ref{Fig5}. A snapshot of the numerical solution of the surface wave transmission problem is shown in Fig. \ref{Fig6}. The numerical scheme is based on that described in the appendix of \cite{sharma2015diffraction} and \cite{Bifurcated}.

\section{Far-field behavior for lattice strip: reflection and transmission coefficient}\label{FarField}
So far we used the mathematical convenience of the presence of a positive imaginary part of the frequency $\upomega$. For the purpose of physical problem, it is useful to look at the limit $\upomega_2\to0^+.$ In this conservative limit, the total energy flux in terms of all outgoing wavemodes can be subdivided into the energy flux ahead of (resp. behind) the surface interface. In other words, one portion of the given energy flux carried by the incident wave mode \citep{brillouin1946wave} is sent back into the side same as that of incidence and the remaining is transmitted into the other side of the surface interface. This applies in general for arbitrary wave modes which are incident and in particular to the surface wave mode incident. However, in this case of surface wave incidence, a splitting is further possible in terms of energy flux carried by the surface wave modes and all other wave modes. Using the dispersion relation,
the group velocity \citep{brillouin1946wave} of the propagating wave with a given wave number $\upxi$ can be determined (see \ref{app_surface_strip}).

Analogous to \eqref{uincwaveS}, let the eigenmodes ahead and behind the interface be denoted by
\begin{equation}\begin{split}
{\hat{a}}_{{A};{\mathtt{y}}}({z}_{{A}}), \quad {\hat{a}}_{{B};{\mathtt{y}}}({z}_{{B}}),
\label{modespn}
\end{split}\end{equation}
respectively.
Now as ${\mathtt{x}}\to+\infty,$
\begin{equation}\begin{split}
{\su}_{{\mathtt{x}}, 0}&=\frac{1}{2\pi i}\oint\limits_{{\mathbb{T}}}{\mathfrak{L}}_+{\mathfrak{C}}_+{z}^{{\mathtt{x}}-1}d{z}= \frac{1}{2\pi i}\oint\limits_{{\mathbb{T}}}\frac{{\mathcal{B}}_+}{{\mathcal{A}}_+}{\mathfrak{C}}_+{z}^{{\mathtt{x}}-1}d{z}\\
&\sim\sum\limits_{\text{Res}}\frac{{\mathcal{B}}_+({z})}{{\mathcal{A}}_+({z})}{\mathfrak{C}}_+({z}){z}^{{\mathtt{x}}-1}\\
&={{\mathrm{A}}{\hat{a}}_{{A};0}({z}_{\inc})}\sum\limits_{{\mathcal{A}}_+({z})=0}\frac{{\mathcal{B}}_+({z})}{{\mathcal{A}}'_+({z})}\bigg(\frac{{\mathfrak{L}}_+^{-1}({z}_{\inc})}{{z}-{z}_{\inc}}\\
&-(\alpha_{{A}}-\alpha_{{B}})\mathcal{U}({z}_{\inc})\sum\limits_{\widetilde{{z}}|{\mathcal{B}}_+(\widetilde{{z}})=0}{\mathcal{P}}({\widetilde{{z}}})\frac{\mathcal{G}_{-}(\widetilde{{z}})}{{z}-\widetilde{{z}}}\bigg){z}^{{\mathtt{x}}}.
\label{discWHNsolp}
\end{split}\end{equation}
Hence, as ${\mathtt{x}}\to+\infty,$
\begin{equation}\begin{split}
{\su}_{{\mathtt{x}}, {\mathtt{y}}}&\sim{{\mathrm{A}}{\hat{a}}_{{A};0}({z}_{\inc})}\sum\limits_{{\mathcal{A}}_+({z})=0}\frac{{\mathcal{B}}_+({z})}{{\mathcal{A}}'_+({z})}\bigg(\frac{{\mathfrak{L}}_+^{-1}({z}_{\inc})}{{z}-{z}_{\inc}}\\
&-(\alpha_{{A}}-\alpha_{{B}})\mathcal{U}({z}_{\inc})\sum\limits_{\widetilde{{z}}|{\mathcal{B}}_+(\widetilde{{z}})=0}{\mathcal{P}}({\widetilde{{z}}})\frac{\mathcal{G}_{-}(\widetilde{{z}})}{{z}-\widetilde{{z}}}\bigg)\frac{{\hat{a}}_{{A};{\mathtt{y}}}({z})}{{\hat{a}}_{{A};0}({z})}{z}^{{\mathtt{x}}}.
\label{farfieldsolp}
\end{split}\end{equation}
Similarly, as ${\mathtt{x}}\to-\infty,$
\begin{equation}\begin{split}
{\su}_{{\mathtt{x}}, 0}&=\frac{1}{2\pi i}\oint\limits_{{\mathbb{T}}}{\mathfrak{L}}_-^{-1}{\mathfrak{C}}_-{z}^{{\mathtt{x}}-1}d{z}= \frac{1}{2\pi i}\oint\limits_{{\mathbb{T}}}\frac{{\mathcal{A}}_-}{{\mathcal{B}}_-}{\mathfrak{C}}_-{z}^{{\mathtt{x}}-1}d{z}\\
&\sim
-\sum\limits_{\text{Res}}\frac{{\mathcal{A}}_-({z})}{{\mathcal{B}}_-({z})}{\mathfrak{C}}_-({z}){z}^{{\mathtt{x}}-1}\\
&=-{\su}^{\inc}_{{\mathtt{x}}, 0}
-{{\mathrm{A}}{\hat{a}}_{{A};0}({z}_{\inc})}\sum\limits_{{\mathcal{B}}_-({z})=0}\frac{{\mathcal{A}}_-({z})}{{\mathcal{B}}'_-({z})}\bigg(-\frac{{\mathfrak{L}}_+^{-1}({z}_{\inc})}{{z}-{z}_{\inc}}\\
&-(\alpha_{{A}}-\alpha_{{B}})\mathcal{U}({z}_{\inc})\sum\limits_{\widetilde{{z}}|{\mathcal{B}}_+(\widetilde{{z}})=0}{\mathcal{P}}({\widetilde{{z}}})\frac{\mathcal{G}_{-}({z})-\mathcal{G}_{-}(\widetilde{{z}})}{{z}-\widetilde{{z}}}\bigg){z}^{{\mathtt{x}}}.
\label{discWHNsoln}
\end{split}\end{equation}
Hence, as ${\mathtt{x}}\to-\infty,$
\begin{equation}\begin{split}
{\su}_{{\mathtt{x}}, {\mathtt{y}}}&\sim-{\su}^{\inc}_{{\mathtt{x}}, {\mathtt{y}}}
-{{\mathrm{A}}{\hat{a}}_{{A};0}({z}_{\inc})}\sum\limits_{{\mathcal{B}}_-({z})=0}\frac{{\mathcal{A}}_-({z})}{{\mathcal{B}}'_-({z})}\bigg(-\frac{{\mathfrak{L}}_+^{-1}({z}_{\inc})}{{z}-{z}_{\inc}}\\
&-(\alpha_{{A}}-\alpha_{{B}})\mathcal{U}({z}_{\inc})\sum\limits_{\widetilde{{z}}|{\mathcal{B}}_+(\widetilde{{z}})=0}{\mathcal{P}}({\widetilde{{z}}})\frac{\mathcal{G}_{-}({z})-\mathcal{G}_{-}(\widetilde{{z}})}{{z}-\widetilde{{z}}}\bigg)\frac{{\hat{a}}_{{B};{\mathtt{y}}}({z})}{{\hat{a}}_{{B};0}({z})}{z}^{{\mathtt{x}}}.
\label{farfieldsoln}
\end{split}\end{equation}

The amplitude of the reflected surface wave, which has the form ${\su}^{{{A}}}_{{\mathtt{x}},0}={{\mathrm{A}}{\hat{a}}_{{A};0}({z}_{{{A}}})}{\sconst}_{{S:}{{A}}}\exp({i{\upxi}_{{A}} {\mathtt{x}}})$, can be obtained from above expression \eqref{discWHNsolp} where ${\upxi}_{{A}}=-{\upxi}_{\inc}$. In fact, ${z}_{{A}}={z}^{-1}_{\inc}$ and ${\mathcal{A}}_+({z}_{{A}})={\mathcal{A}}_+({z}^{-1}_{\inc})=0,$ so that the surface wave reflection coefficient is given by
\begin{equation}\begin{split}
{\sconst}_{{S:}{{A}}}&=\frac{{\hat{a}}_{{A};0}({z}_{\inc})}{{\hat{a}}_{{A};0}({z}_{{{A}}})}\frac{{\mathcal{B}}_+({z}_{{A}})}{{\mathcal{A}}'_+({z}_{{A}})}
\bigg(\frac{{\mathfrak{L}}_+^{-1}({z}_{\inc})}{{z}_{{A}}-{z}_{\inc}}-(\alpha_{{A}}-\alpha_{{B}})\mathcal{U}({z}_{\inc})\sum\limits_{\widetilde{{z}}|{\mathcal{B}}_+(\widetilde{{z}})=0}{\mathcal{P}}({\widetilde{{z}}})\frac{\mathcal{G}_{-}(\widetilde{{z}})}{{z}_{{A}}-\widetilde{{z}}}\bigg).
\label{refcoefft}
\end{split}\end{equation}
The amplitude of the transmitted surface wave of the form ${\su}^{{{B}}}_{{\mathtt{x}},0}={{\mathrm{A}}{\hat{a}}_{{B};0}({z}_{{{B}}})}{\sconst}_{{S:}{{B}}}\exp({i{\upxi}_{{B}} {\mathtt{x}}})$ can be obtained from expression \eqref{discWHNsoln}. Indeed, ${\mathcal{B}}_-({z}_{{{B}}})=0,$ so that the surface wave transmission coefficient is given by
\begin{equation}\begin{split}
{\sconst}_{{S:}{{B}}}&=\frac{{\hat{a}}_{{A};0}({z}_{\inc})}{{\hat{a}}_{{B};0}({z}_{{{B}}})}\frac{{\mathcal{A}}_-({z}_{{{B}}})}{{\mathcal{B}}'_-({z}_{{{B}}})}\bigg(\frac{{\mathfrak{L}}_+^{-1}({z}_{\inc})}{{z}_{{{B}}}-{z}_{\inc}}+(\alpha_{{A}}-\alpha_{{B}})\mathcal{U}({z}_{\inc})\sum\limits_{\widetilde{{z}}|{\mathcal{B}}_+(\widetilde{{z}})=0}{\mathcal{P}}({\widetilde{{z}}})\frac{\mathcal{G}_{-}({z}_{{{B}}})-\mathcal{G}_{-}(\widetilde{{z}})}{{z}_{{{B}}}-\widetilde{{z}}}\bigg).
\label{transcoefft}
\end{split}\end{equation}

\section{Energy leaked away from the surface}\label{Leakage}

\begin{figure}[htb!]
\center{\includegraphics[width=.9\textwidth]{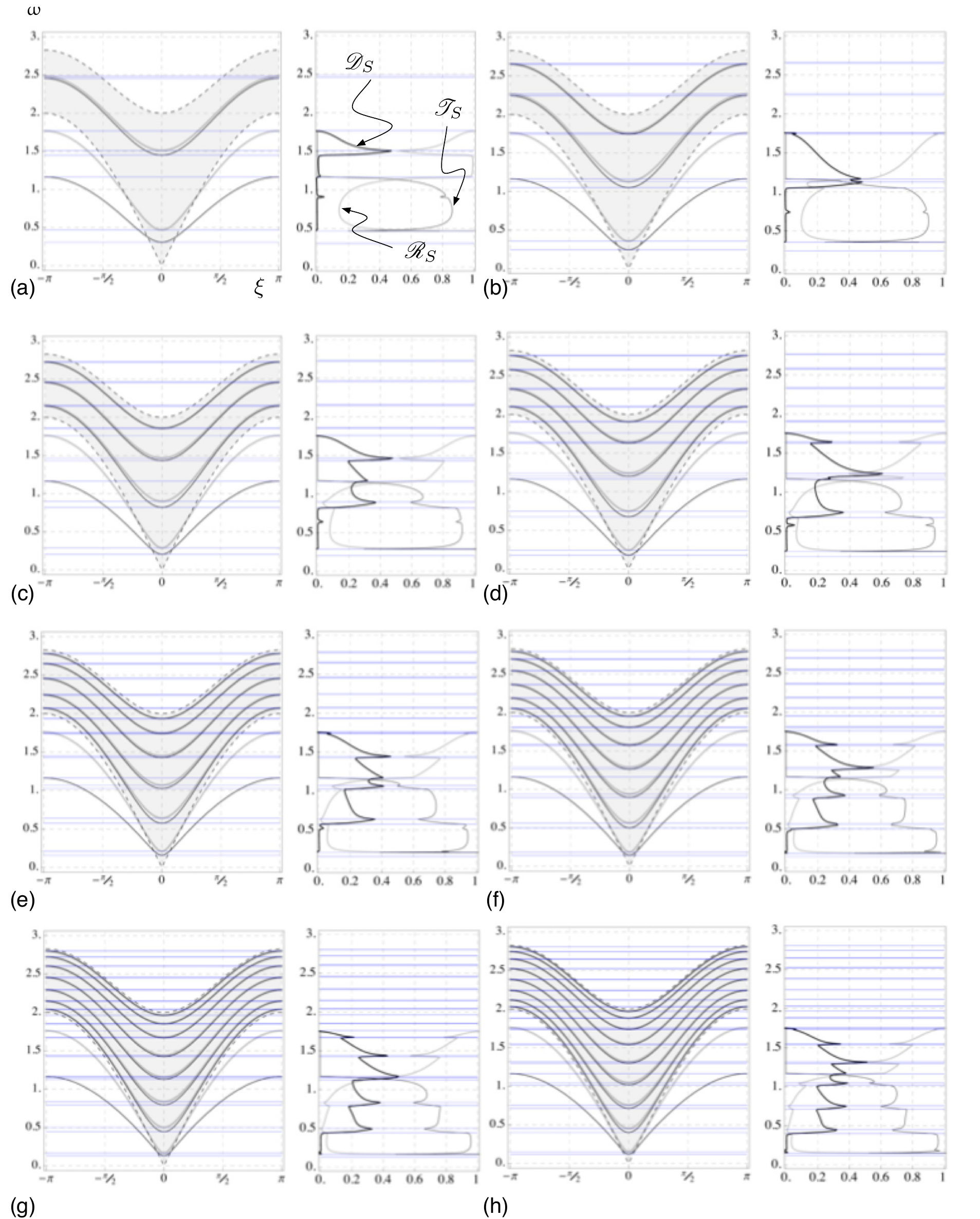}}
\caption[ ]{Illustration of the reflectance ${\mathscr{R}}_{S}$, transmittance ${\mathscr{T}}_{S}$, and the leaked energy flux ${\mathscr{D}}_{S}$.
(a)--(h) correspond to $\mathtt{N}=2, 3, \dotsc, 8$. In all plots, $\alpha_{{A}}=1.4$, $\alpha_{{B}}=1.5$, $m_{{A}}=2$, $m_{{A}}=5$.
The vertical axis represents $\upomega$. The horizontal axis on left plot for each part is the wave number $\upxi$. The horizontal axis on right plot for each part is ${\mathscr{R}}_{S}, {\mathscr{T}}_{S}, {\mathscr{D}}_{S}$. }
\label{Fig7}
\end{figure}

\begin{figure}[ht!]
\center{\includegraphics[width=\textwidth]{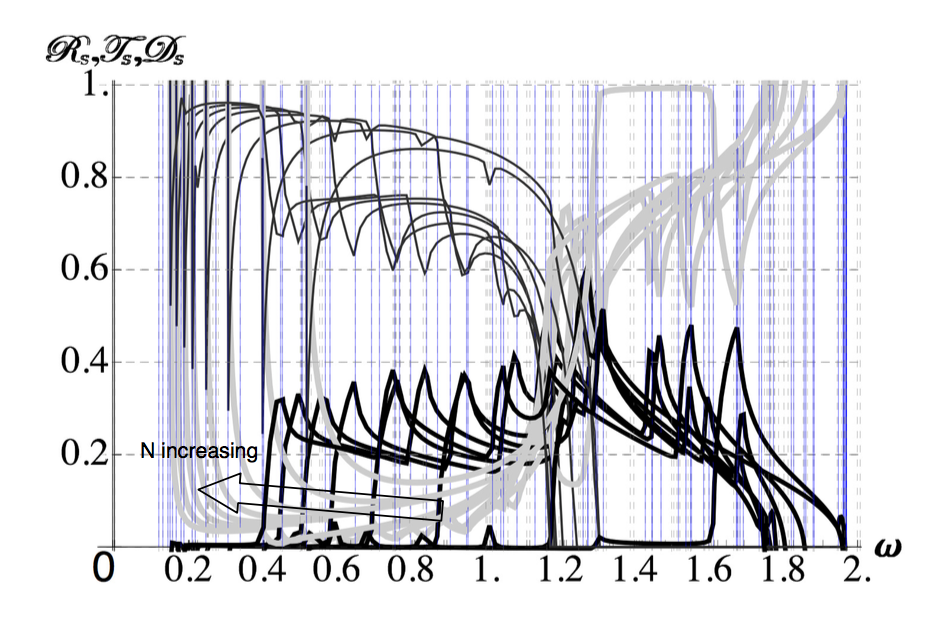}}
\caption[ ]{Illustration of the reflectance (black thin curves), transmittance (gray thick curves), and the leaked energy flux (black thick curves) by superposition of the eight values of $\mathtt{N}$; $\mathtt{N}=2, 3, \dotsc, 8$ with $\alpha_{{A}}=1.4$, $\alpha_{{B}}=1.5$, $m_{{A}}=2$, $m_{{A}}=5$, as shown in Fig. \ref{Fig7}.}
\label{Fig8}
\end{figure}

Following the standard procedure to calculated the energy fluxes \citep{brillouin1946wave,achenbach2012wave}, using
\eqref{uincwaveS} and \eqref{energyfluxS} we get
for the incident surface wave
\begin{equation}\begin{split}
\mathcal{E}_{\inc}&=\frac{1}{2}{\upomega}^2|{\mathtt{V}_g}({z}_{\inc})||{\mathrm{A}}|^2,
\end{split}\end{equation}
while for the reflected surface wave
\begin{equation}\begin{split}
\mathcal{E}_{{{A}}}&=\frac{1}{2}{\upomega}^2|{\mathtt{V}_g}({z}_{{A}})||{\mathrm{A}}{\sconst}_{{S:}{{A}}}|^2,
\end{split}\end{equation}
and finally for the transmitted surface wave
\begin{equation}\begin{split}
\mathcal{E}_{{{B}}}&=\frac{1}{2}{\upomega}^2|{\mathtt{V}_g}({z}_{{B}})||{\mathrm{A}}{\sconst}_{{S:}{{B}}}|^2,
\end{split}\end{equation}
where the group velocity is determined by \eqref{surfacewavegroup} and \eqref{surfacewavegroupY} and $\mathcal{M}({z})$
is given by \eqref{normconst}.
In terms of the expression \eqref{refcoefft}, since ${\mathtt{V}_g}({z}_{\inc})=-{\mathtt{V}_g}({z}_{{A}}), \mathcal{M}({z}_{\inc})=\mathcal{M}({z}_{{A}})$, the surface reflectance is given by
\begin{equation}\begin{split}
{\mathscr{R}}_{S}=\frac{\mathcal{E}_{{{A}}}}{\mathcal{E}_{\inc}}=|{\sconst}_{{S:}{{A}}}|^2.
\label{reflance}
\end{split}\end{equation}
On the other hand and in terms of the expression \eqref{transcoefft}, the surface transmittance is given by
\begin{equation}\begin{split}
{\mathscr{T}}_{S}=\frac{\mathcal{E}_{{{B}}}}{\mathcal{E}_{\inc}}=|{\sconst}_{{S:}{B}}|^2|\dfrac{{\mathtt{V}_g}({z}_{{B}})}{{\mathtt{V}_g}({z}_{\inc})}|.
\label{transance}
\end{split}\end{equation}
Using the expressions \eqref{reflance} and \eqref{transance}, the energy lost into the bulk, per unit energy flux of the incident surface wave, during the process of transmission on the surface is given by
\begin{equation}\begin{split}
{\mathscr{D}}_{S}&=1-{\mathscr{R}}_{S}-{\mathscr{T}}_{S}=1-|{\sconst}_{{S:}{{A}}}|^2-|{\sconst}_{{S:}{B}}|^2|\dfrac{{\mathtt{V}_g}({z}_{{B}})}{{\mathtt{V}_g}({z}_{\inc})}|,
\label{dissance}
\end{split}\end{equation}
where \eqref{refcoefft} and \eqref{transcoefft} need to be employed.
By inspection it is clear that
\eqref{refcoefft}
as well as \eqref{transcoefft} hold for any other wave mode in the portions behind and ahead the interface, respectively, i.e., ${\sconst}_{{A}}'$ and ${\sconst}_{{B}}'$. Hence, by the conservation of energy it follows that
\begin{equation}\begin{split}
{\mathscr{D}}_{S}&=\sum_{{z}_{{A}}'\ne{z}_{{A}}}|{\sconst}_{{A}}'|^2|\dfrac{{\mathtt{V}_g}({z}_{{A}}')}{{\mathtt{V}_g}({z}_{\inc})}|+\sum_{{z}_{{B}}'\ne{z}_{{B}}}|{\sconst}_{{B}}'|^2|\dfrac{{\mathtt{V}_g}({z}_{{B}}')}{{\mathtt{V}_g}({z}_{\inc})}|.
\label{dissance2}
\end{split}\end{equation}
An illustration of the physically important entity ${\mathscr{D}}_{S}$ relative to the incident wave frequency $\upomega$ is presented in Fig. \ref{Fig7}. When the frequency is such that there is no other wave modes in both lattice strips, besides the surface wave modes, then according to \eqref{dissance2} it is expected that ${\mathscr{D}}_{S}=0.$ A superposition of the energy flux plots from Fig. \ref{Fig7} has been provided in Fig. \ref{Fig8}.

\section{Physical implications and inferences: a discussion}
Surface waves, \textit{i.e.} waves whose amplitude decays exponentially with distance form the surface, play an important role as they carry an information about the material properties and microstructure of the media. So they are widely used in such filed as nondestructive evaluation. On the other hand, surface waves are also used in numerous acousto-electronic devices such as filters and transformers. Miniaturization of the acousto-electronic devices with current nanotechnologies leads to necessity to take into account surface stresses and energy. Here within the lattice dynamics  we discuss the shear wave propagation across a surface interface keeping the material in the bulk homogeneous. As the considered here square lattice with surface discrete interface is a discrete counterpart of the linear Gurtin--Murdoch model, we can say that obtained results relates also to this continuum limit. Nevertheless the rigorous consideration of the continuum limit requires further analysis. 

Considering Figs.~\ref{Fig7} and \ref{Fig8} we can see now that the surface interface can significantly change the total picture of wave transmittance, reflectance and leakage into the bulk depending on frequency $\upomega$. This make possible to control the wave propagation for a given frequency changing the material properties. From the physical point of view such interface can treated as grain or subgrain boundary or as an interface between surface areas subjected different treatment. So a step-wise inhomogeneity only may dramatically change the wave propagation. Further consideration of a system of interfaces may change surface properties with respect to wave propagation, as in the case of so-called metasurfaces, see reviews by  \cite{holloway2012overview,chen2016review}. For example, a system of parallel interfaces is quite similar to hyperbolic metasurfaces as in \citep{ji2014broadband,high2015visible,li2018infrared} but this interference  should be also rigorously considered in the forthcoming papers.

\section{Concluding remarks}
Following \cite{Victor_Bls_surf1} here we present the exact solution for anti-plane surface wave propagation within the lattice dynamics. Considering surface linear interface as a boundary between material particles of two types which both are different from the ones in the bulk we analyze the wave transmission and reflection along the interface, as well as energy leakage ${\mathscr{D}}_{S}$ into the bulk. In other words, we consider the anti-plane waves in a half-space which boundary consists of two parts with different surface properties. We have provided a closed form expression for the flux ${\mathscr{D}}_{S}$ for a lattice strip. Analogous expression can be also found for the lattice half space however that is not attempted here as large enough width of the strip is deemed to be a very good approximation for the half space (remember we are dealing with incident surface wave which is exponentially decaying into the depth of half space).
The presented here technique shows high efficiency for modelling of surface-related phenomena. 
The presented analysis also applies to the bulk wave incidence for half space as well as any other lattice wave mode besides surface wave for lattice strip.

\section*{Acknowledgments}
B.L.S. acknowledges the support of SERB MATRICS grant MTR/2017/000013.
V.A.E. acknowledges the support   by grant 14.Z50.31.0036 awarded to R. E. Alekseev Nizhny Novgorod Technical University by Department of Education and Science of the Russian Federation.

\section*{References}
\bibliographystyle{model2-names}\biboptions{authoryear}
\bibliography{./interfacelatticeantiplane}

\appendix
\section{Surface wave in a strip clamped on one side}
\label{app_surface_strip}
Consider a propagating wave along positive ${\mathtt{x}}$ direction on a strip with fixed boundary at ${\mathtt{y}}=-\mathtt{N}$ and free boundary at ${\mathtt{y}}=0$ (see any of the left or right portions of the half space shown in Fig. \ref{Fig3}). Following \cite{sharma2017linear}, let the wave form be described by
\begin{equation}\begin{split}
{\hat{a}}_{{\mathtt{y}}}e^{i{\upxi}{\mathtt{x}}}e^{-i{\upomega} t}.
\label{surfwavegen}
\end{split}\end{equation}
We have
\eqref{eqmotion} and \eqref{eqmotionBCa} (replace $A$ by $S$). By \eqref{latticestrip2},
\begin{equation}\begin{split}
{\hat{a}}_{{\mathtt{y}}}={\hat{a}}_{0}\frac{{\lambda}^{-{\mathtt{y}}-\mathtt{N}}-{\lambda}^{{\mathtt{y}}+\mathtt{N}}}{{\lambda}^{-\mathtt{N}}-{\lambda}^{\mathtt{N}}}={\hat{a}}_{0}\frac{{\mathtt{U}}_{{\mathtt{y}}+\mathtt{N}-1}}{{\mathtt{U}}_{\mathtt{N}-1}}, \quad{\vartheta}={\tfrac{1}{2}}{\mathtt{Q}},
\label{eigenmode}
\end{split}\end{equation}
where
(recall \eqref{defQ})
${\mathtt{Q}}={\lambda}+{\lambda}^{-1}=4-{z}-{z}^{-1}-{\upomega}^2.$

Further (recall \eqref{deffreq}),
\begin{equation}\begin{split}
-m_{S}{\upomega}^2{\hat{a}}_{0}&
=\alpha_{S}(2\cos{\upxi}-2){\hat{a}}_{0}+(\frac{{\lambda}^{1-\mathtt{N}}-{\lambda}^{-1+\mathtt{N}}}{{\lambda}^{-\mathtt{N}}-{\lambda}^{\mathtt{N}}}-1){\hat{a}}_{0},
\end{split}\end{equation}
which requires for non-trivial solutions
\begin{equation}\begin{split}
-m_{S}{\upomega}^2&=\alpha_{S}(2\cos{\upxi}-2)+(\frac{{\lambda}^{1-\mathtt{N}}-{\lambda}^{-1+\mathtt{N}}}{{\lambda}^{-\mathtt{N}}-{\lambda}^{\mathtt{N}}}-1).
\end{split}\end{equation}
In term of the Chebyshev polynomials \citep{Mason} and using the recent analysis provided by \citep{sharma2017linear}, inspection of
\eqref{discWHF1}
reveals that above equation is precisely ${\mathfrak{K}}_{S}=0,$ i.e.,
\begin{equation}\begin{split}
{\mathtt{U}}_{\mathtt{N}-2}+F_{S}{\mathtt{U}}_{\mathtt{N}-1}=0,
\label{surfacewavedispersion}
\end{split}\end{equation}
as $F_{S}({z})=m_{S}{\upomega}^2-1+\alpha_{S}({z}+{z}^{-1}-2),$ with ${z}=e^{-i{\upxi}}.$

The group velocity ${\mathtt{V}_g}$ of the wave modes, in particular the surface wave, is easily found too. Let ${z}=e^{-i {\upxi}}$
Taking \eqref{surfacewavedispersion} as the implicit definition of ${\upomega}({\upxi})$, a derivative w.r.t. $\xi$ (note $\frac{\partial}{\partial{\upxi}}{\vartheta}={\vartheta}'=-{\tfrac{1}{2}}(-i{z}+i{z}^{-1})-{\upomega}{\upomega}'$) leads to
\begin{equation}\begin{split}
(-{\tfrac{1}{2}}(-i{z}+i{z}^{-1})-{\upomega}{\upomega}')\frac{1}{{\vartheta}^2-1}(({\mathtt{N}-1}){\mathtt{T}}_{\mathtt{N}-1}+{\mathtt{N}}F_{S}{\mathtt{T}}_{\mathtt{N}})\\
+(2m_{S}{\upomega}{\upomega}'+\alpha_{S}(-i{z}+i{z}^{-1})){\mathtt{U}}_{\mathtt{N}-1}=0.
\end{split}\end{equation}
Using \eqref{surfacewavedispersion1} and \eqref{surfacewavedispersion2},
finally, the group velocity is given by
\begin{equation}\begin{split}
{\mathtt{V}_g}({z})={\mathtt{V}_g}(e^{-i{\upxi}})\equiv{\mathtt{V}_g}({\upxi})={\upomega}'=\frac{i}{2{\upomega}}({z}-{z}^{-1})\frac{Y_{S}-2\alpha_{S}}{Y_{S}-2m_{S}},
\label{surfacewavegroup}
\end{split}\end{equation}
where
\begin{equation}\begin{split}
Y_{S}=\frac{1}{{\vartheta}^2-1}(\mathtt{N}-1+(2{\mathtt{N}}-1){{\vartheta}}F_{S}+{\mathtt{N}}F_{S}^2).
\label{surfacewavegroupY}
\end{split}\end{equation}

The component ${\hat{a}}_{0}$ can be written in a form so that the wave form is normalized. The analysis based on \eqref{eigenmode} yields
$\sum_{{\mathtt{y}}=0}^{-\mathtt{N}+1}{\hat{a}}_{{\mathtt{y}}}^2={\hat{a}}_{0}^2\mathcal{M}$,
where
\begin{equation}\begin{split}
\mathcal{M}({z})={\mathtt{U}}^{-2}_{\mathtt{N}-1}\sum\limits_{{\mathtt{y}}=0}^{-\mathtt{N}+1}{\mathtt{U}}^2_{{\mathtt{y}}+\mathtt{N}-1}.
\label{normconst}
\end{split}\end{equation}
In above,
\begin{equation}\begin{split}
\sum\limits_{{\mathtt{y}}=0}^{-\mathtt{N}+1}{\mathtt{U}}^2_{{\mathtt{y}}+\mathtt{N}-1}
=\sin^{-2}{\upeta}(\frac{1}{4}+{\tfrac{1}{2}} \mathtt{N}-\frac{1}{4}\frac{\sin(1+2\mathtt{N}){\upeta}}{\sin{\upeta}}).
\end{split}\end{equation}
Though, it is not necessary for our purpose most of the time, for calculation of energy flux it is needed to normalize.
In a general case, i.e., including the surface wave, the energy density can be found. The kinetic energy at a given ${\mathtt{x}}$ is given by
\begin{equation}\begin{split}
&{\tfrac{1}{2}}{\upomega}^2(m_{S}{\hat{a}}_{0}^2+\sum_{{\mathtt{y}}=-1}^{-\mathtt{N}+1}{\hat{a}}_{{\mathtt{y}}}^2)={\tfrac{1}{2}}{\upomega}^2((m_{S}-1){\hat{a}}_{0}^2+\sum_{{\mathtt{y}}=0}^{-\mathtt{N}+1}{\hat{a}}_{{\mathtt{y}}}^2)\\
&={\tfrac{1}{2}}{\upomega}^2{\hat{a}}_{0}^2((m_{S}-1)+\mathcal{M}({z}_{S})).
\end{split}\end{equation}
The energy flux (in the direction of the group velocity) via a surface wave with wave form ${\su}^{S}_{{\mathtt{x}},0}={\mathrm{A}}_0{\hat{a}}_{0}\exp({i{\upxi}_{S} {\mathtt{x}}})$ at ${\mathtt{y}}=0$ is given by
\begin{equation}\begin{split}
\mathcal{E}_{S}&=\frac{1}{2}{\upomega}^2|{\mathtt{V}_g}({z}_{S})||{\mathrm{A}}_0|^2(m_{S}{\hat{a}}_{0}^2+\sum_{{\mathtt{y}}=-1}^{-\mathtt{N}+1}{\hat{a}}_{{\mathtt{y}}}^2)\\
&=\frac{1}{2}{\upomega}^2|{\mathtt{V}_g}({z}_{S})||{\mathrm{A}}_0|^2,
\label{energyfluxS}
\end{split}\end{equation}
where the normalization is achieved with
\begin{equation}\begin{split}
{\hat{a}}_{0}({z})=\dfrac{1}{\sqrt{\mathcal{N}({z}_{S})}},
\end{split}\end{equation}
where
\begin{equation}\begin{split}
\mathcal{N}({z}_{S})=m_{S}-1+\mathcal{M}({z}_{S}), \quad \mathcal{M}({z})=\frac{\frac{1}{4}+{\tfrac{1}{2}} \mathtt{N}-\frac{1}{4}{\mathtt{U}}_{2\mathtt{N}}}{(1-{\vartheta}^2){\mathtt{U}}^2_{\mathtt{N}-1}}.
\end{split}\end{equation}
For $m_{S}=1$, above relations reduce to that for a uniform lattice strip \cite{brillouin1946wave,Bifurcated}.

The surface wave band lies within the limits ${\upomega}_{\min}$ and ${\upomega}_{\max}$ found as follows. With ${z}=1$, $2{\vartheta}=2-{\upomega}^2, F_{S}=m_{S}{\upomega}^2-1$ where ${\upomega}$ is given by
\begin{equation}\begin{split}
{\mathtt{U}}_{\mathtt{N}-2}({\vartheta})+(m_{S}{\upomega}^2-1){\mathtt{U}}_{\mathtt{N}-1}({\vartheta})=0
\end{split}\end{equation}
With ${z}=-1$, $2{\vartheta}=6-{\upomega}^2, F_{S}=m_{S}{\upomega}^2-1-4\alpha_{S}$ where ${\upomega}$ is given by
\begin{equation}\begin{split}
{\mathtt{U}}_{\mathtt{N}-2}({\vartheta})+(m_{S}{\upomega}^2-1-4\alpha_{S}){\mathtt{U}}_{\mathtt{N}-1}({\vartheta})=0.
\end{split}\end{equation}

Also \eqref{surfacewavedispersion} can be written as
$-{\mathtt{U}}_{\mathtt{N}}+({\mathtt{Q}}+F_{S}){\mathtt{U}}_{\mathtt{N}-1}=0, $
and by adding $-2{\mathtt{T}}_{\mathtt{N}}+({\mathtt{Q}}+2F_{S}){\mathtt{U}}_{\mathtt{N}-1}=0$, i.e.,
\begin{equation}\begin{split}
2{\mathtt{T}}_{\mathtt{N}}=({\mathtt{Q}}+2F_{S}){\mathtt{U}}_{\mathtt{N}-1}.
\label{surfacewavedispersion1}
\end{split}\end{equation}

Also \eqref{surfacewavedispersion} can be written as
$(1+F_{S}{\mathtt{Q}}){\mathtt{U}}_{\mathtt{N}-2}-F_{S}{\mathtt{U}}_{\mathtt{N}-3}=0, $
and by adding $(2+F_{S}{\mathtt{Q}}){\mathtt{U}}_{\mathtt{N}-2}+2F_{S}{\mathtt{T}}_{\mathtt{N}-1}=0$, i.e.,
\begin{equation}\begin{split}
2{\mathtt{T}}_{\mathtt{N}-1}=-({\mathtt{Q}}+2F_{S}^{-1}){\mathtt{U}}_{\mathtt{N}-2}=(F_{S}{\mathtt{Q}}+2){\mathtt{U}}_{\mathtt{N}-1}.
\label{surfacewavedispersion2}
\end{split}\end{equation}

\section{Auxiliary derivation}
\label{app_u00}
We recall that $\su_{{\mathtt{x}}, 0}=\frac{1}{2\pi i}\int\nolimits_{{\mathcal{C}}}{\su}^F{z}^{{\mathtt{x}}-1}d{z}$. Motivated by the technique to obtain the expression for the site facing the rigid constraint, as demonstrated in \cite{sharma2015diffraction2,BlsNearRigid},
\begin{equation}\begin{split}
\su_{0, 0}=\frac{1}{2\pi i}\int\limits_{{\mathcal{C}}}{\su}^F{z}^{-1}d{z}.
\end{split}\end{equation}
Deforming the contour ${\mathcal{C}}$ to ${\mathcal{C}}_\infty$ (with radius ${\mathit{R}}_\infty$ tending to infinity), with ${\mathfrak{C}}_-(0)$ given by \eqref{expnCn0},
\begin{equation}\begin{split}
\su_{0, 0}&=\frac{1}{2\pi i}\int\limits_{{\mathcal{C}}_\infty}{\su}^F{z}^{-1}d{z}
-{{\mathrm{A}}{\hat{a}}_{{A};0}({z}_{\inc})}+\sum\limits_{{z}|{\mathcal{B}}_-({z})=0}\dfrac{{{\mathrm{A}}{\hat{a}}_{{A};0}({z}_{\inc})}}{{\mathfrak{L}}_+({z}_{\inc})}\dfrac{{\mathcal{A}}_-({z})}{{\mathcal{B}}'_-({z})}\dfrac{1}{{z}-{z}_{\inc}}\\
&+(\alpha_{{A}}-\alpha_{{B}})\su^{\totwave}_{0, 0}\sum\limits_{\widetilde{{z}}|{\mathcal{B}}_+(\widetilde{{z}})=0}{\mathcal{P}}({\widetilde{{z}}})\sum\limits_{{z}|{\mathcal{B}}_-({z})=0}\dfrac{1}{{z}-\widetilde{{z}}}(\dfrac{G_{\mathtt{N};-}({{z}})}{{\mathcal{B}}'_-({z})}+\mathcal{G}_{-}(\widetilde{{z}})\dfrac{-{\mathcal{A}}_-({z})}{{\mathcal{B}}'_-({z})}),
\end{split}\end{equation}
where the first term is zero, so that
\begin{equation}\begin{split}
\su^{\totwave}_{0, 0}&=\dfrac{{{\mathrm{A}}{\hat{a}}_{{A};0}({z}_{\inc})}}{{\mathfrak{L}}_+({z}_{\inc})}\dfrac{\sum\limits_{{z}|{\mathcal{B}}_-({z})=0}\dfrac{{\mathcal{A}}_-({z})}{{\mathcal{B}}'_-({z})}\dfrac{1}{{z}-{z}_{\inc}}}{1-(\alpha_{{A}}-\alpha_{{B}})\sum\limits_{\widetilde{{z}}|{\mathcal{B}}_+(\widetilde{{z}})=0}\sum\limits_{{z}|{\mathcal{B}}_-({z})=0}\dfrac{{\mathcal{P}}({\widetilde{{z}}})}{{z}-\widetilde{{z}}}\dfrac{G_{\mathtt{N};-}({{z}})-\mathcal{G}_{-}(\widetilde{{z}}){\mathcal{A}}_-({z})}{{\mathcal{B}}'_-({z})}}.
\end{split}\end{equation}

\end{document}